\documentclass[useAMS,usenatbib]{mn2e}

\usepackage{times}
\usepackage{graphicx}
\usepackage{rotating}

\def\pcm3{{\rm\thinspace cm^{-3}}}

\def\n_h{{\rm n_{H}}}

\def\NH1{{$N_{\rm HI}~$}}

\def\ga{{\rm\thinspace gauss}}

\def\Msun{\hbox{$\rm\thinspace M_{\odot}$}}

\def\approxlt{\mathrel{\hbox{\rlap{\lower .5ex \hbox {$\sim$}}
        \raise .15 ex \hbox{$<$}}}}
\def\approxgt{\mathrel{\hbox{\rlap{\lower .5ex \hbox {$\sim$}}
        \raise .15 ex \hbox{$>$}}}}

\def\la{\mathrel{\hbox{\rlap{\hbox{\lower4pt\hbox{$\sim$}}}\hbox{$<$}}}}
\def\ga{\mathrel{\hbox{\rlap{\hbox{\lower4pt\hbox{$\sim$}}}\hbox{$>$}}}}
\newbox\grsign \setbox\grsign=\hbox{$>$} \newdimen\grdimen
\grdimen=\ht\grsign
\newbox\simlessbox \newbox\simgreatbox \newbox\simpropbox
\setbox\simgreatbox=\hbox{\raise.5ex\hbox{$>$}\llap
     {\lower.5ex\hbox{$\sim$}}}\ht1=\grdimen\dp1=0pt
\setbox\simlessbox=\hbox{\raise.5ex\hbox{$<$}\llap
     {\lower.5ex\hbox{$\sim$}}}\ht2=\grdimen\dp2=0pt
\setbox\simpropbox=\hbox{\raise.5ex\hbox{$\propto$}\llap
     {\lower.5ex\hbox{$\sim$}}}\ht2=\grdimen\dp2=0pt
\def\simgreat{\mathrel{\copy\simgreatbox}}
\def\simless{\mathrel{\copy\simlessbox}}

\title[Proper motion L and T dwarf candidate members of the Pleiades]{Proper motion L and T dwarf candidate members of the Pleiades}
\author[S. L. Casewell et al.]{S. L. Casewell$^{1}$\thanks{E-mail:
slc25@star.le.ac.uk}, P. D. Dobbie$^{1,2}$, S. T. Hodgkin$^{3}$, E. Moraux$^{4}$, R. F. Jameson$^{1}$, 
 \newauthor N. C. Hambly$^{5}$, J. Irwin$^{3}$ and N. Lodieu$^{6,1}$\\ 
$^{1}$Department of Physics and Astronomy, University of Leicester, University Road, Leicester LE1 7RH, UK\\
$^{2}$Anglo-Australian Observatory, PO Box 296, Epping NSW 1710 Australia\\
$^{3}$CASU, Institute of Astronomy,University of Cambridge, Maddingley Road, Cambridge, CB3 0HA, UK \\
$^{4}$Laboratoire d'Astrophysique, Observatoire de Grenoble, Universit\'e Joseph Fourier, BP 53, 38041 Grenoble Cedex 9, France\\
$^{5}$Scottish Universities Physics Alliance (SUPA),
     Institute for Astronomy, School of Physics, University of Edinburgh, \\
     Royal Observatory, Blackford Hill, Edinburgh EH9 3HJ \\
$^{6}$ Instituto de Astrof\'isica de Canarias, V\'ia L\'actea s/n, E-38205 La Laguna, Tenerife, Spain\\}

\begin{document}

\date{January 2007 }

\pagerange{\pageref{firstpage}--\pageref{lastpage}} \pubyear{2007}

\maketitle

\label{firstpage}

\begin{abstract}
We present the results of a deep optical-near-infrared multi-epoch survey covering 2.5 square degrees of the Pleiades open star cluster to
search for new very-low-mass brown dwarf members. A significant ($\sim 5$ year) epoch difference exists between the optical (CFH12k \textit{I}-,
\textit{Z}-band) and near infrared (UKIRT WFCAM \textit{J}-band) observations. We construct \textit{I},\textit{I}-\textit{Z} and \textit{Z},\textit{Z}-\textit{J} colour magnitude diagrams to select candidate
cluster members. Proper motions are computed for all candidate members and compared to the background field objects to further refine the
sample. We recover all known cluster members within the area of our survey. In addition, we have discovered 9 new candidate brown dwarf cluster
members. The 7 faintest candidates have red \textit{Z}-\textit{J} colours and show blue near-infrared colours.
These are consistent with being L and T-type Pleiads. Theoretical models
predict their masses to be around 11M$_{\rm Jup}$.

\end{abstract}


\section{Introduction}

The initial mass spectrum (IMS), the number of objects manufactured per unit mass interval, is 
an outcome of the star formation process which can be constrained via observation. Consequently,
empirical determinations of the form of the IMS can be used to critically examine our theoretical 
understanding of the complexities of star formation. In recent years there has been a particular 
emphasis on building a solid comprehension of the mechanisms by which very-low-mass stars, brown 
dwarfs and free-floating planetary mass objects form (e.g. Boss 2001; Bate 2004; Goodwin, Whitworth \& Ward-Thompson 2004; Whitworth \& Goodwin 2005).
Nevertheless, one key question which remains unanswered is what is the lowest possible mass of object 
that can be manufactured by the star formation process$?$ From a theoretical stance, traditional models
predict that if substellar objects form like stars, via the fragmentation and collapse of molecular clouds, 
then there is a strict lower mass limit to their manufacture of 0.007-0.010 $\Msun$ (Padoan \& Nordlund, 2002). This is set by the 
rate at which the gas can radiate away the heat released by the compression (e.g. Low \& Lynden-Bell, 
1976). However, in more elaborate theories, 
magnetic fields could cause rebounds in collapsing 
cloud cores which might lead to the decompressional cooling of the primordial gas, a lowering of the Jeans mass 
and hence the production of gravitationally bound fragments with masses of only $\sim$0.001 $\Msun$ (Boss, 2001). 
In contrast, if feedback from putative winds and outflows driven by the onset of deuterium burning play a 
role, the smallest objects which form via the star formation process may be restricted to masses equal to 
or greater than the deuterium burning limit ($\sim$0.013 $\Msun$; Adams \& Fatuzzo, 1996). 

Recent work on very young clusters ($\tau$$<$10 Myrs) and star formation regions e.g. $\sigma$-Orionis,
the Trapezium, IC348 and Upper Sco (B\'ejar et al., 2001; Muench et al., 2002; Muench et al., 2003; Lodieu et al., 2007a)
 suggests that the initial mass function continues slowly rising
down to masses of the order M$\sim$0.01 M$_{\odot}$, at least in these environments. Indeed, it has been
claimed that an object with a mass as low as 2-3 M$_{\rm Jup}$ has been unearthed in $\sigma$-Ori (Zapatero-Osorio et al, 2002).
However, the cluster membership of $\sigma$-Ori 70 is disputed by Burgasser et al. (2004). Furthermore, 
mass estimates for such young substellar objects derived by comparing their observed properties to the
predictions of theoretical evolutionary tracks remain somewhat controversial. Baraffe et al. (2002)
have shown that to robustly model the effective temperature and luminosity of a low mass object with
an age less than $\sim$1 Myr, evolutionary calculations need to be coupled to detailed simulations
of the collapse and accretion phase of star formation. As the current generation of evolutionary models
start from arbitrary initial conditions, theoretical predictions for ages less than a few Myrs must
be treated with a fair degree of caution. Indeed, the few available dynamical mass measurements of pre-main 
sequence objects indicate that models tend to underestimate mass by a few tens of percent in the range 
0.3$\simless$M$\simless$1.0 M$_{\odot}$ (see Hillenbrand \& White, 2004 for review). A recent dynamical mass
 measurement of the 50-125 Myrs old object AB Dor C (spectral type $\sim$M8), the first for a pre-main 
sequence object with M$<$0.3 M$_{\odot}$, suggests that the discrepancy between model predictions and 
reality might be even larger at lower masses, with the former underestimating mass by a factor 2-3 at 
M$\sim$0.1 M$_{\odot}$ (Close et al., 2005). However, this conclusion is dependent on the assumed age of 
AB Dor, which is currently a matter of great contention (Luhman, Stauffer \& Mamajek, 2005; Janson et al., 2006). On 
the positive side, Zapatero-Osorio et al., (2004) have determined the masses of the brown dwarf binary components 
of GJ 569 Bab and their luminosities and effective temperatures are in agreement with theoretical predictions, for an age of 
300 Myr. More recently, Stassun, Mathieu \& Valenti (2006) discuss an eclipsing brown dwarf binary in the Orion nebula star forming 
region and find the large radii predicted by theory for a very young dwarf. Surprisingly, they find that the secondary is hotter
than the more massive primary. Clearly further work is still needed to support the predictions of theoretical models.

It is clearly important to search for the lowest mass objects, not only in the young clusters, but also in more mature clusters, such 
as the Pleiades.
 The results of previous 
surveys of the Pleiades indicate that the present day cluster mass function, across the stellar/substellar 
boundary and down to M$\sim$0.02 M$_{\odot}$ (based on the evolutionary models of the Lyon Group), 
can be represented by a slowly rising power law model, dN/dM$\propto$M$^{-\alpha}$. For example, from their 
Canada-France-Hawaii Telescope (CFHT) survey conducted at \textit{R} and \textit{I} and covering  2.5 sq. degrees, Bouvier et al.
(1998) identified 17 candidate brown dwarfs (\textit{I}$_{\rm C}$$\ge$17.8) and derived a power law index of $\alpha$=0.6.
From their 1.1 sq degrees Isaac Newton Telescope (INT) survey conducted at \textit{I} and \textit{Z}, with follow-up work undertaken 
at \textit{K}, Dobbie et al. (2002) unearthed 16 candidate substellar members and found a power law of index $\alpha$=0.8 
to be compatible with their data. Jameson et al. (2002) showed that 
a powerlaw of index $\alpha$=0.41$\pm$0.08 was consistent with the observed mass function over the range 0.3$
\simgreat$M$\simgreat$0.035 M$_{\odot}$. This study used a sample of 49 probable brown dwarf members assembled from the four 
most extensive CCD surveys of the cluster available at the time, the International Time Project survey 
(Zapatero Osorio et al., 1998), the CFHT survey (Bouvier et al., 1998; Moraux, Bouvier \& Stauffer, 2001), the Burrell Schmidt 
survey (Pinfield et al., 2000) and the INT survey (Dobbie et al., 2002). The CFHT survey was subsequently extended to an area of 6.4 sq. degrees 
(at \textit{I} and \textit{Z}) and unearthed a total of 40 candidate brown dwarfs. Moraux et al. (2003) applied statistical arguments
to account for non-members in their sample and derived a power law index of $\alpha$=0.6. Most recently, Bihain 
et al. (2006) have used deep \textit{R}, \textit{I}, \textit{J} and \textit{K} band photometry and proper motion measurements to unearth 6 robust L type 
Pleiades members in an area of 1.8 sq. degrees with masses in the range 0.04-0.02 M$_{\odot}$ and derived a power law index of $\alpha$=0.5$\pm$0.2.

Here we report the results of a new optical/infrared survey of 2.5 sq. degrees of the Pleiades, the aim of which 
is to extend empirical constraints on the cluster mass function down to the planetary mass regime (M$\sim
$0.01 M$_{\odot}$). In the next section we describe the observations acquired/used as part for this study, their
reduction, their calibration and their photometric completeness. In subsequent sections we describe how we have
identified candidate brown dwarf members on the basis of colours and proper motions. We use our new results to constrain 
the form of the cluster mass function and conclude by briefly discussing our findings in the context of star formation 
models.

\section{Observations, Data Reduction and Survey Completeness}

\subsection{The \textit{J} band imaging and its reduction}

Approximately 3.0 square degrees of the Pleiades cluster was observed in the \textit{J} band using the Wide Field Camera (WFCAM) on 
the United Kingdom Infrared Telescope (UKIRT) between the dates of 29/09/2005 and 08/01/2006. WFCAM is a near infrared imager 
consisting of 4 Rockwell Hawaii-II (HgCdTe 2048x2048) arrays with 0.4'' pixels, arranged such that 4 separate pointings 
(pawprints) can be tiled together to cover a 0.75 sq. degree region of sky (see http://www.ukidss.org/technical/technical.html for diagram). 
A total of four tiles
were observed in a mixture of photometric and non-photometric conditions but in seeing of typically $\approx$ 1.0 arcsecond or 
better. To ensure that the images were properly sampled we employed the 2$\times$2 microstep mode. The locations on the sky of 
our tiles (shown in Figure 1) were chosen to provide maximum overlap with the optical fields surveyed in 2000
by the Canada-France-Hawaii telescope and CFH12k camera but also to avoid bright stars and areas of significant interstellar extinction.

\begin{figure*}
\begin{center}
\scalebox{0.725}{\includegraphics[angle=270]{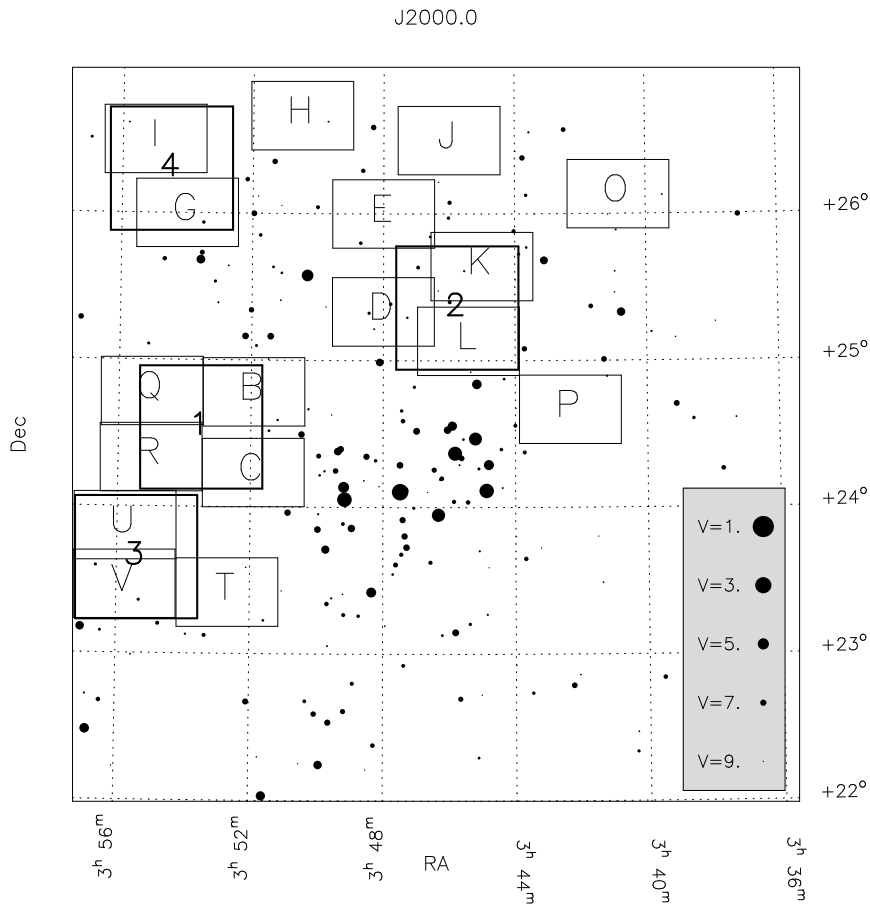}}
\caption{The regions imaged at \textit{I}, \textit{Z} and \textit{J} with the CFHT and UKIRT. The CFH12k pointings (light rectangular 
outlines) are labelled alphabetically as in Moraux et al. (2003), while the WFCAM tiles (bold square outlines) are labelled 
numerically, ranging from 1 to 4. Note that the observations avoid the region of high reddening to the south of the Merope and
the bright stars in vicinity of the cluster centre.}
\end{center}
\end{figure*}

The images were reduced at the Cambridge Astronomical Survey Unit (CASU) using procedures which have been custom written 
for the treatment of WFCAM data. In brief, each frame was 
debiased, dark corrected and then flat fielded. The individual dithered images were stacked before having an object detection 
routine run on them. The detection procedure employs a core radius of 5 pixels, and identifies objects as islands of more than 4 
interconnected pixels with flux $>$1.5$\sigma$ above the background level.
The frames were astrometrically calibrated using point sources in the
Two micron All Sky Survey (2MASS) catalogue. These solutions, in general,
had a scatter of less than 0.1 arcseconds. The photometric calibration
employed by the CASU pipeline also relies on 2MASS data (there are typically hundreds of 2MASS calibrators per detector) and is found to be
accurate to $\approx$2\% in good conditions (see Warren et al., 2007, Hodgkin et al., 2007 for details).

In measuring our photometry we used an aperture of 2", which is
approximately twice the core radius of point sources. This 2'' diameter of the
aperture is also twice the seeing FWHM.
 The reduction
pipeline also attempts to classify each source depending on its
morphology (e.g. galaxy, star, noise). However, at the limit of the data
this classification becomes less reliable. Therefore, in our subsequent
analysis we chose to define as stellar all objects which lie within 3 sigma of the
stellar locus, where sigma is defined according to Irwin et al. (in
prep).

\subsection{The far-red optical imaging and a new reduction}

As part of this work we have used a subset (2.54 square degrees) of the far-red optical data obtained in the course of the \textit{IZ} survey of the Pleiades 
conducted in 2000 by Moraux et al. (2003). The relevant CFH12k data were extracted from the Canadian Astrophysical Data Center archive
and were reprocessed at Cambridge University using the CASU optical imaging pipeline (Irwin \& Lewis, 2001). In brief, these data were
 bias 
subtracted and corrected for non-linearity prior to flat fielding. Fringe maps, which were constructed for each photometric band 
from images obtained during the observing run, were used to remove the effects of interference between night sky lines in the CCD 
substrate. Subsequently, sources at a level of significant of 3$\sigma$ or greater were morphologically classified and aperture photometry 
obtained for each.  A World Coordinate System (WCS) was determined for each frame by cross-correlating these sources with the Automated 
Plate Measuring (APM) machine catalogue (Irwin, 1985). The approximately 100 common objects per CCD chip lead to an internal accuracy of 
typically better than 0.3 ''. 
The photometry was calibrated onto a CFH12k \textit{I} and \textit{Z} natural system using stars with near zero colour 
(\textit{B}-\textit{V}-\textit{R}-\textit{I}$\approx$0) in Landolt standard field SA98 (Landolt, 1992) which was observed the same nights as the 
science data. The systematic errors in the photometry were calculated by comparing the photometry of overlapping fields as in Moraux et al. (2003).
The photometry was found to be accurate to $\approx$3\%.

\subsection{The completeness of datasets}

To estimate the completeness of our IR images, we injected fake stars with magnitudes in the range \textit{J}=12-22 into
each of the 16 chips of every WFCAM frame and re-ran the object detection software with the same parameters that
were used to detect the real sources. To avoid significantly increasing the density of all sources in the data we inserted
only 200 fake stars per chip in a given run. To provide meaningful statistics we repeated this whole procedure ten times.
Subsequently, we calculated percentage completeness at a given magnitude by
taking the ratio of the number of fake stars recovered to the number of fake stars injected into a given
magnitude bin (and multiplying by 100). We note that a 100\% recovery rate was never achieved at any magnitude
since a small proportion of the fake stars always fell sufficiently close to other sources to be overlooked by
the object detection algorithm. This method was also applied to determine the completeness of the \textit{I} and 
\textit{Z} band CFH12k data. However, the magnitude range of the fake stars was adjusted to be consistent with the
different saturation and faint end magnitude limits of these data. The results of this procedure for all 3
photometric bands are shown in Table 1.

A glance at this table indicates that the IR data are in general 90\% complete to \textit{J}$\approx$19.7, although Field
3 is slightly less deep, due to moonlight and poor seeing. In this case the proximity of the moon led to higher background counts. The \textit{I} data are
typically 90\% and 50\% complete to \textit{I}=22.5 and 23.5 respectively. The corresponding completeness limits for the \textit{Z} band
data are \textit{Z}=21.5 and 22.5 respectively.

 \begin{table*}
\caption{50 and 90\% completeness figures for the optical and infrared fields. The positioning of these fields is shown in Figure 1. Note that while 
WFCAM field 1 corresponds to CFHT fields B, C, R and Q, the individual pawprints, do not correspond on a one to one basis - i.e. field1\_00 does not correspond to 
field B.}
\begin{center}
\begin{tabular}{l c c c c c c c c c}
\hline
Field name& \multicolumn {2}{|c|}{\textit{I}}&\multicolumn {2}{|c|}{\textit{Z}}&WFCAM tile name&WFCAM pawprint name&\multicolumn {2}{|c|}{\textit{J}}\\
&50\%&90\%&50\%&90\%&&&50\%&90\%\\
\hline
B&23.2&22.5&22.3&21.5&field1& 00&20.9&19.9\\
C&23.7&22.6&22.6&21.6&field1& 01&20.9&20.1\\
R&24.0&23.0&22.9&21.6&field1& 10&20.9&19.8\\
Q&23.7&22.5&22.7&21.6&field1& 11&20.9&19.7\\
K&23.6&22.5&23.0&21.9&field2& 00&20.9&19.7\\
L&24.0&22.7&23.0&21.8&field2& 01&20.9&19.9\\
D&23.7&22.4&23.0&21.7&field2& 10&21.0&19.9\\
&&&&&field2& 11&20.9&19.7\\
U&23.5&22.5&22.9&21.7&field3& 00&19.5&18.8\\
V&23.8&22.5&22.7&21.7&field3& 01&19.0&17.7\\
T&23.6&22.5&22.6&21.5&field3& 10&19.6&18.6\\
&&&&&field3& 11&18.9&17.7\\
I&23.9&22.3&23.1&22.0&field4& 00&20.8&19.7\\
G&23.7&22.7&23.4&22.3&field4& 01&20.8&19.7\\
&&&&&field4& 10&20.8&19.7\\
&&&&&field4& 11&20.8&19.7\\
\hline
\end{tabular}
\end{center}
\end{table*}

\section{Analysis of the data}

\subsection{Photometric selection of candidate cluster members}

An initial photometrically culled sample of candidate brown dwarfs has
been obtained from the \textit{I},\textit{I}-\textit{Z} colour-magnitude diagram (Figure 2) where the
120 Myr NEXTGEN (Baraffe et al., 1998) and DUSTY (Chabrier et al., 2000)
model isochrones (modified to take into account the Pleiades distance of
134 pc e.g. Percival, Salaris \& Groenewegen, 2005) served as a guide to the location of the
Pleiades sequence. With the uncertainties in both the photometry and the
age of the cluster in mind, we selected all objects classed as stellar in both the \textit{I} and \textit{Z} data, 
 which in the magnitude
range 16.5 $<$ \textit{I} $<$ 22.5 lay no more than 0.25 magnitudes to the left of the DUSTY
isochrone. All the candidate Pleiads found by Moraux et al. (2003) and Bihain et al. (2006) lay within $\pm$0.25 magnitudes of the DUSTY model.
 Thus our selection criterion is 0.25 magnitudes to the left of the DUSTY model. Below \textit{I}=22.5, the DUSTY model is not red enough to account for known field stars, and so is inappropriate in this 
effective temperature regime. We have calculated an approximate
field star sequence from Tinney, Burgasser \& Kirkpatrick (2003) and Hawley et al (2002) and lowered it by 2 magnitudes. This results in the line 
 \textit{I}-\textit{Z}$=$ (\textit{I}-19.0)/3.5. 
This selection is conservative, and is particularly aimed at removing the bulk of the red tail of the background stars.
 Subsequently,
the initial list of candidates was cross-correlated with our J band
photometric catalogue (using a matching radius of 2 arcseconds) and a
refined photometrically culled sample obtained using the \textit{Z},\textit{Z}-\textit{J}
colour-magnitude diagram (Figure 3). These objects are also shown on the \textit{J}, \textit{I}-\textit{J} colour-magnitude diagram (Figure 4). 
As before, the 120 Myr model isochrones were used as a guide to the
location of the cluster sequence. With the photometric uncertainties in
mind, all candidates with \textit{Z}$\leq$20 were retained. All candidates with  20$<$\textit{Z}$<$21 and \textit{Z}-\textit{J}$\geq$1.6 were also retained. Finally, all candidates with \textit{Z}$>$21 and \textit{Z}-\textit{J}$\geq$1.9 were retained. These constraints are conservative and are based on 
the field L and T  dwarfs sequence (\textit{Z}-\textit{J}$\geq$3, Chiu et al., 2006) since the DUSTY models are known to be inappropriate in this 
effective temperature regime.  Since
our survey is limited by the depth of the \textit{I} band data, all candidates with
Z$>$20 and no \textit{I} band counterpart were also kept.
\begin{figure*}
\begin{center}
\scalebox{0.50}{\includegraphics[angle=270]{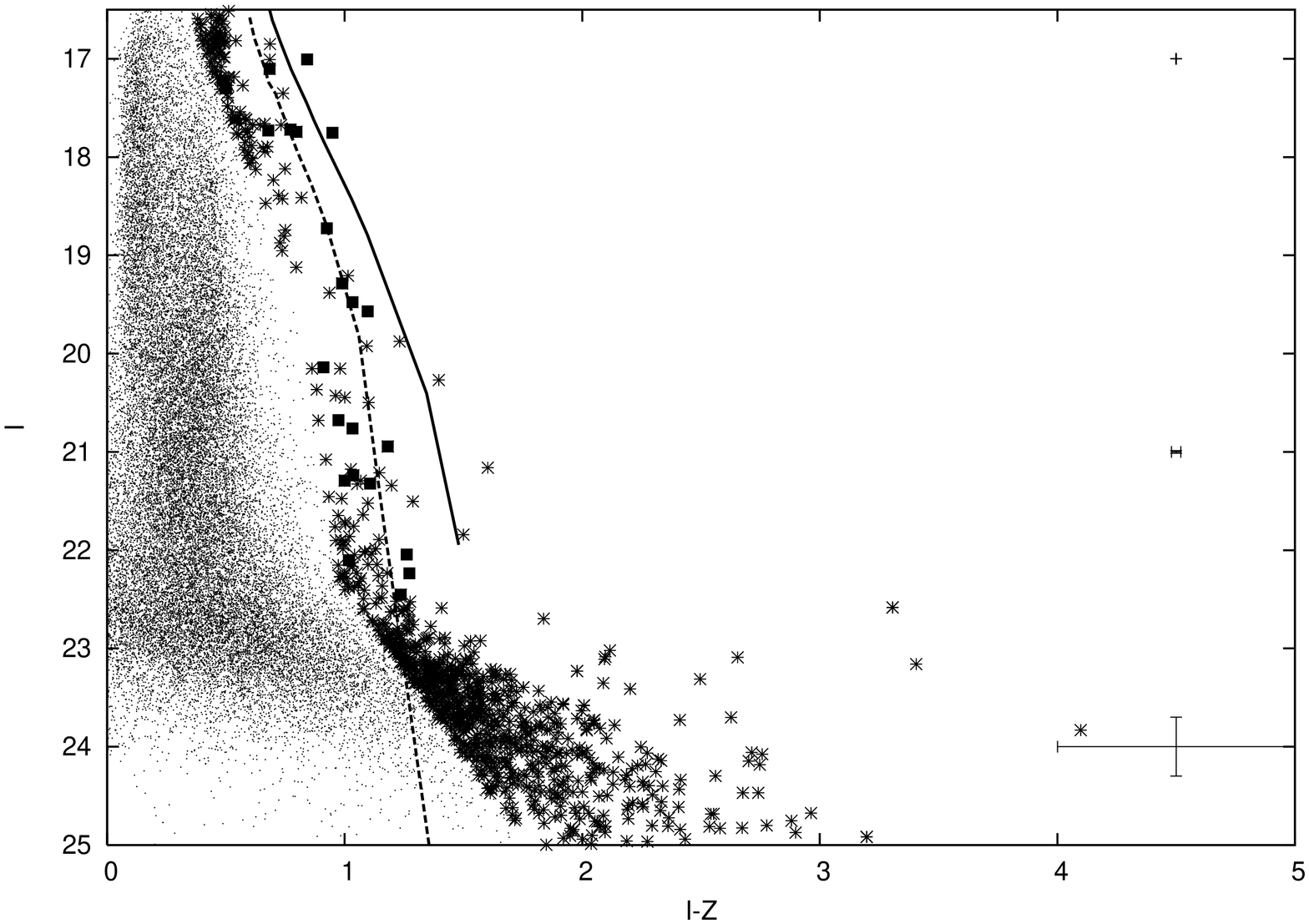}}
\caption{The \textit{I},\textit{I}-\textit{\textit{Z}} CMD for the whole of field 1. The solid line is the NEXTGEN model, and the dotted line the DUSTY model. 
The small points are all objects that were classed as stellar in both \textit{I} and \textit{Z} data.
The crosses are all objects that met the following selection criteria: classed as stellar in both \textit{I} and \textit{Z} data, 
for 16.5 $<$ \textit{I} $<$ 22.5, they must lie no more than 0.25 magnitudes to the left of the DUSTY
isochrone, for \textit{I}$\geq$22.5, they must lie to the right of the line,
 \textit{I}-\textit{Z}$=$ (\textit{I}-19.0)/3.5.  The filled squares are the previously identified cluster candidate members from Bihain et al. (2006),
Moraux et al. (2003) and Bouvier et al. (1998), plotted to highlight the cluster sequence.
}
\end{center}
\end{figure*}
\begin{figure*}
\begin{center}
\scalebox{0.50}{\includegraphics[angle=270]{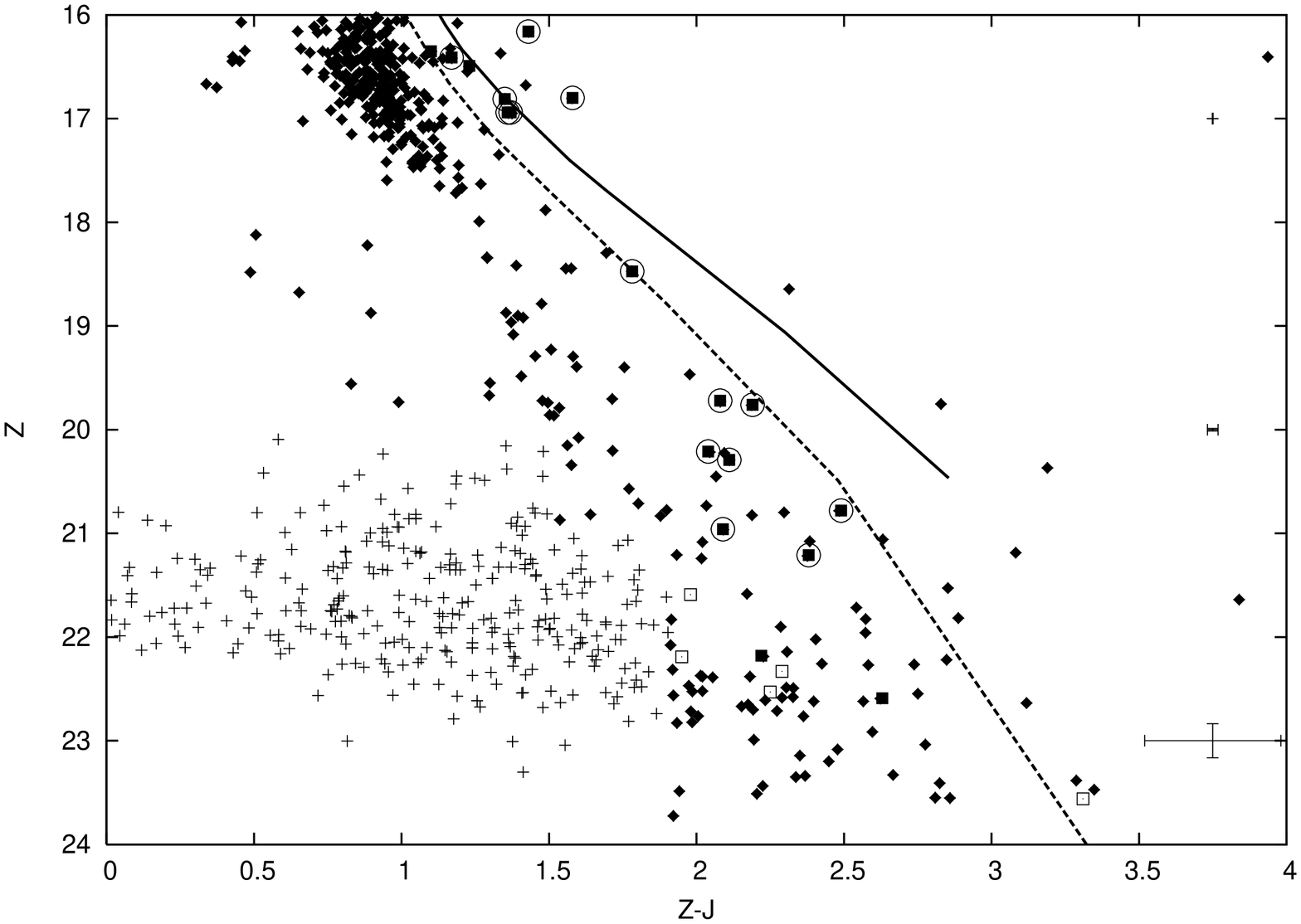}}
\caption{The \textit{Z},\textit{Z}-\textit{J} CMD for the whole of the survey. The solid line is the NEXTGEN model, and the dotted line the DUSTY model. 
The crosses are all the objects selected from the \textit{I},\textit{I}-\textit{Z} (crosses on Figure 2.). The filled diamonds are all objects 
that met our selection criteria from the \textit{I},\textit{I}-\textit{Z},and \textit{Z}, \textit{Z}-\textit{J} CMDs. 
These were selected for proper motion analysis, and were found to be non members. The filled
squares are our candidate cluster members (objects that remained after proper motion analysis).
The squares are our \textit{ZJ} only candidates for all four fields that remained after proper motion analysis. The previously identified probable members from Bihain et al. (2006),
Moraux et al. (2003) and Bouvier et al. (1998) that remained after our proper motion analysis are identified by open circles around the plotted symbols.
}
\end{center}
\end{figure*}
\begin{figure*}
\begin{center}
\scalebox{0.50}{\includegraphics[angle=270]{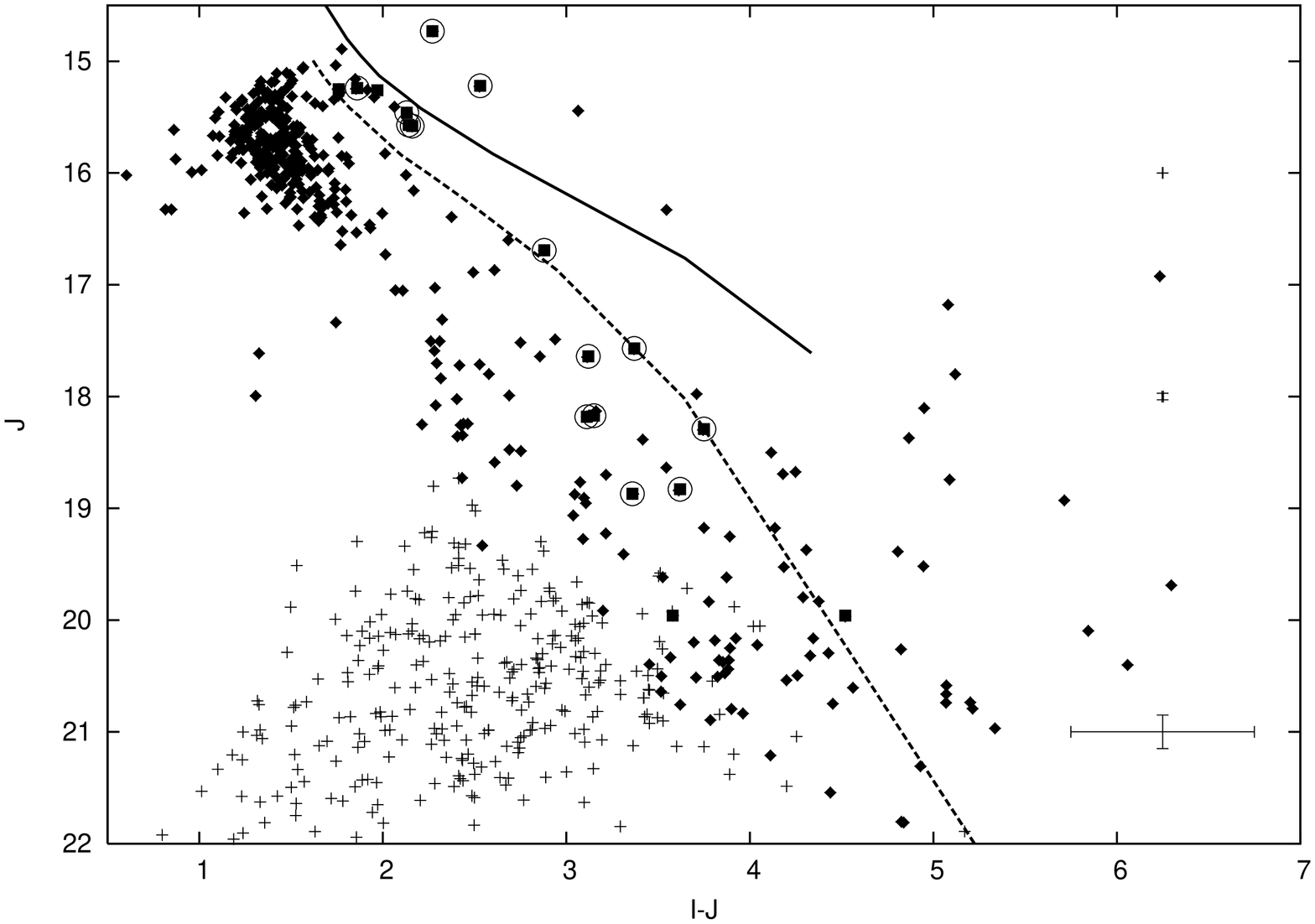}}
\caption{The \textit{J},\textit{I}-\textit{J} CMD for the whole of the survey. The solid line is the NEXTGEN model, and the dotted line the DUSY model. 
The crosses are all the objects selected from the \textit{I},\textit{I}-\textit{Z} (crosses on Figure 2.). The filled diamonds are all objects 
that met our selection criteria from the \textit{I},\textit{I}-\textit{Z},and \textit{Z}, \textit{Z}-\textit{J} CMDs. 
These were selected for proper motion analysis, and were found to be non members. The filled
squares are our candidate cluster members (objects that remained after proper motion analysis).
The previously identified members from Bihain et al. (2006),
Moraux et al. (2003) and Bouvier et al. (1998) that remained after our proper motion analysis are identified by open circles around the plotted symbols.
}
\end{center}
\end{figure*}
\subsection{Refining the sample using astrometric measurements}

To weed out non-members we have measured the proper motion of each candidate brown dwarf, using the \textit{Z} and \textit{J} band data where the 
epoch difference was 5 years. In this process, only 
objects lying within 2 arcminutes of each candidate were chosen as potential astrometric reference stars. This 
compromise provided a sufficiently large number of sources but at the same time minimised the effects of non-linear 
distortions in the images. Furthermore, objects with large ellipticity ($>$0.2), classed by the photometric 
pipeline as non-stellar in the \textit{Z} band data  and with \textit{Z}$<$16 or 
\textit{Z}$>$20 were rejected. This ensured that, in the main, the astrometric reference sources were not of very low S/N in the 
\textit{J} band or saturated in the optical data. These criteria generally provided at least 20 suitable stars per candidate
brown dwarf. 

Six coefficient transforms between the epoch 1 \textit{Z} band images and the epoch 2 \textit{J} band images were calculated using 
routines drawn from the STARLINK SLALIB package. The iterative fitting rejects objects having residuals greater
than 3$\sigma$, where $\sigma$ is robustly calculated as the median of
absolute deviation of the reference star residuals, scaled by the
appropriate factor (1.48) to yield an equivalent RMS. Once the routine had converged the relative proper motions in pixels were calculated by 
dividing the fitting residuals of each candidate by the epoch difference. For our data the epoch difference is 
approximately 5 years. Subsequently, the astrometric motion in milliarcseconds per year in RA and DEC was derived by 
folding these values through the World Coordinate System transform matrix of the relevant WFCAM image.

To estimate the errors on our proper motions measurements, we have injected fake stars
into both the \textit{Z} and \textit{J} band data, in a similar fashion to that described in section 2.3.
However, here we have determined the difference between the inserted
position and the
photometric pipeline estimate of the centroid of each star. Assuming that
the differences
between these two locations are normally distributed, we have divided the
fake stars into
3 magnitude bins in each photometric band (\textit{Z}$\leq$21, 23$\geq$\textit{Z}$>$21, 24$\geq$\textit{Z}$>$23, 21$\geq$\textit{J}$>$17)
and fit 2d Gaussians to estimate the 1-sigma centroiding uncertainty as a function of source brightness.

We find that in the \textit{Z} band data, for objects with magnitudes \textit{Z}$\leq$21, the
centroiding uncertainty
is equivalent to 3 mas yr$^{-1}$ in each axis, while for objects with
23$\geq$\textit{Z}$>$21 this number
increases to 8 mas yr$^{-1}$. For our faintest \textit{Z} band objects, 24$\geq$\textit{Z}$>$23,
the centroiding
uncertainty is equivalent to 12 mas yr$^{-1}$ in each axis.  In the \textit{J}
band data, for objects
with magnitudes 21$\geq$\textit{J}$>$17, the centroiding uncertainty is equivalent to 5
mas yr$^{-1}$ in each axis. 
Thus for our brightest candidates (\textit{Z}$<$21, \textit{J}$<$19), the
quadratic sum of the \textit{Z}
and \textit{J} band centroiding errors is less than or comparable to the RMS of the
residuals of the
linear transform fit, which is typically 5-10 mas yr$^{-1}$ in each axis.
We adopt this latter
quantity as the proper motion uncertainty in both the RA and DEC
directions for these objects.
It is worth noting at this point that both the stars and brown dwarfs of
the Pleiades appear to
be in a state of dynamical relation (e.g. Pinfield et al. 1998, Jameson et
al. 2002), where the
velocity dispersion of the members is proportional to 1/M$^{0.5}$, where M
is mass. Based on an
extrapolation of the data in Figure 4 of Pinfield et al. (1998), we would
expect our lowest mass
brown dwarf members (0.01-0.02M$_{\odot}$) to have velocity dispersion of $\sim$ 7
mas yr$^{-1}$. This velocity dispersion should be added quadratically to the above uncertainties. 
Our final adopted proper motion selection, effectively a 
radius of 14 mas yr$^{-1}$ , is described below, and the velocity dispersion is small compared to this.

We fitted the proper motions of all of our photometric candidates (excluding the \textit{ZJ} only candidates) with a 2D Gaussian, which centred around 1.1, -7 mas yr$^{-1}$. This Gaussian had a $\sigma$ 
of 14.0. 
We were not able to fit two Gaussians, one to the background stars and one to the Pleiades dwarfs, as described in Moraux et al.(2003),
 since only $\approx$ 30 objects have the 
correct proper motion for cluster membership. 
Consequently, we only selected objects to be proper motion members if they had proper motions that fell within 1$\sigma$ of the proper motion of
 the cluster at +20.0, -40.0 mas yr$^{-1}$ (Jones 1981; Hambly, Jameson \& Hawkins, 1991; Moraux et al., 2001).  We required the selection criteria to be 1$\sigma$, as extending this to 
2$\sigma$, would seriously overlap with the field stars centred on 0,0.
We did however extend the selection criteria to 1.5$\sigma$, which yielded 14 additional objects, however all were rejected due to their bright, but blue (\textit{I}$\approx$17.0, \textit{I}-\textit{Z}$<$1.0) positions on the \textit{I},\textit{I}-\textit{Z} CMD, which led us to believe that they were field objects.
We also attempted to tighten our selection criteria to a circle with radius 10 mas yr$^{-1}$. This selection meant that we lost as possible members objects PLZJ 78, 9, 77, 23 (see Table 4). PLIZ 79, 9 and 77 have all been identified and confirmed as proper motion members by Bihain et al. (2006), Moraux et al. (2001), and Bouvier et al. (1998). 
 Unfortunately, as we cannot fit two Gaussians to our data, we cannot
 calculate a probability of membership for 
these objects by the standard method as defined by Sanders (1971). 
The proper motion measurements may be found in Table 4, as well as the \textit{I}, \textit{Z}, \textit{J}, \textit{H}  and \textit{K} magnitudes for these candidate members to the cluster.

We have attempted to use control data to determine the level of contamination within our data, however, the numbers involved are very small, 
so any calculated probability will be rather uncertain. 
We used as controls, two circles of radius 14.0 mas yr$^{-1}$, at the same distance from 0,0 proper motion as the Pleiades.
We then separated the data into one magnitude bins, and calculated the probability for each magnitude bin, using equation 1.
\begin{equation}
P_{\rm membership}=\frac{N_{\rm cluster}-N_{\rm control}}{N_{\rm cluster}}
\label{eqno1}
\end{equation}
Where P$_{\rm membership}$ is the probability of membership for that magnitude bin, N$_{\rm cluster}$ is the number of stars and contaminants within the cluster circle in 
that magnitude bin.  N$_{\rm control}$ is the number of dwarfs in the control circle of proper motion space, see Figure 4. N$_{\rm cluster}$ - $N_{\rm control}$
is the number of Pleiads.  It can be seen that the probability depends on where the control circle is located.  Thus as well as using control circles, 
we use an annulus and scale down the count to an area equal to that of a control circle.  Note that Figure 5 is for all of the magnitude bins together. Figure 6 is 
the same as Figure 5, but for the \textit{ZJ} selected objects only.
  The statistics are much poorer for the individual magnitude bins and the probabilities are correspondingly more uncertain. It can be seen in Figure 4 that there is not a symmetrical distribution of proper motions. In fact the distribution in the Vector point diagram,
is a classical "velocity ellipsoid" displaced from zero by reflex motion
from the Sun's peculiar velocity, and happens to be
in the direction of the Pleiades proper motion vector. We have therefore probably underestimated the contamination, as the annulus method of calculating probabilities assumes that the vector point diagram has a circularly symmetric distribution of objects.  
 These probabilities are shown 
in Table 2, and probabilities derived in the same way but for the \textit{ZJ} only candidates can be found in Table 3.

\begin{figure}
\begin{center}
\scalebox{0.35}{\includegraphics[angle=270]{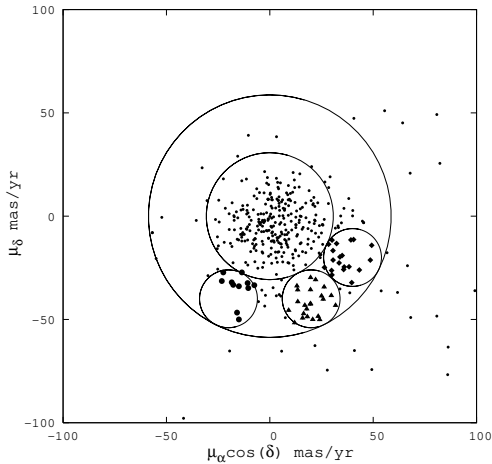}}
\caption{\label{pm}Proper motion vector diagram of the photometrically selected candidate members. The filled triangles are candidate and known 
cluster members.
The filled diamonds and filled circles are the two separate control clusters used. The annulus used for the radial method is also plotted.
}
\end{center}
\end{figure}
\begin{figure}
\begin{center}
\scalebox{0.35}{\includegraphics[angle=270]{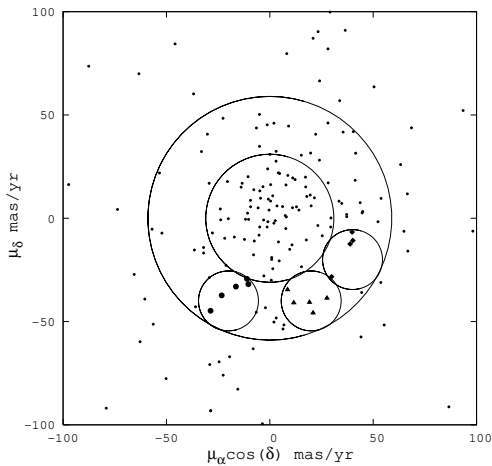}}
\caption{\label{pm}Proper motion vector diagram of the photometrically selected candidate members. The filled triangles are candidate  
cluster members selected from the \textit{Z},\textit{Z}-\textit{J} CMD only.
The filled diamonds and filled circles are the two separate control clusters used. The annulus used for the radial method is also plotted.
}
\end{center}
\end{figure}
An alternative approach to estimating the contamination is the use the field L and T dwarf luminosity functions. Chabrier (2005) gives the T dwarf luminosity function as being 10$^{-3}$ dwarfs/pc$^{3}$/unit $\textit{J}$ mag interval.  Our 7 L and T dwarf candidates cover a total of 0.7 mag in the \textit{J} band. Note PLZJ 323 and 23 may be late L dwarfs but we include them in this analysis.  The volume of space we use is 836 pc$^{2}$, based
on 2.5 square degrees and a distance to the Pleiades of 134$\pm$30 pc (Percival et al., 2005). This distance range corresponds to a distance modulus
range of $\pm$0.5 magnitudes, which is generous, given that the sequence shown in figure 8 is clearly narrower than $\pm$0.5 magnitudes. Thus the 
expected number of contaminating field dwarfs is 0.6. In addition to this, field T dwarfs are unlikely to have the same proper motion as the Pleiades, thus reducing the 0.6 further. For the field L dwarfs with M$_{\rm J}$$\approx$13.0 (i.e. \textit{J}$\approx$18.5 at the distance of the Pleiades) the luminosity function is 3$\times$10$^{-4}$ dwarfs/pc$^{3}$/unit $\textit{J}$ mag interval (Chabrier, 2005). A similar calculation then gives 0.25 contaminating L dwarfs which should be further reduced by considering proper motions. It is thus clear that the field luminosity function indicates that contamination by field L and T dwarfs should be negligible.

 \begin{table*}
\caption{Probability of membership, magnitude range for our methods of calculating probabilities of membership using the annulus as well as the 
two control areas.
}
\begin{center}
\begin{tabular}{l c c c }
\hline
Probability & Probability & Probability & Magnitude range\\
annulus&$\mu_{\alpha}cos \delta$=-20 mas yr$^{-1}$ $\mu_{\delta}$=-40 mas yr$^{-1}$  &$\mu_{\alpha}cos \delta$=+40 mas yr$^{-1}$ $\mu_{\delta}$=-20 mas yr$^{-1}$ &\textit{Z}\\ 
\hline
     0.67&  0.25&0.0& 16 -  17     \\
     0.82&  0.66&0.0& 17 -  18     \\
     0.88&  1.00&0.0& 18 -  19     \\
     0.84&  1.00&0.0& 19 -  20     \\
     1.00&  1.00&1.00& 20 -  21     \\
     0.88&  0.50&1.00& 21 -  22     \\
     0.61&  1.00&0.00& 22 -  23     \\
    
\hline
\end{tabular}
\end{center}
\end{table*}

 \begin{table*}
\caption{Probability of membership, magnitude range for our methods of calculating probabilities of membership using the annulus as well as the 
two control areas for our candidates selected from the \textit{ZJ} data only.
}
\begin{center}
\begin{tabular}{l c c c }
\hline
Probability & Probability & Probability & Magnitude range\\
annulus&$\mu_{\alpha}cos \delta$=-20 mas yr$^{-1}$ $\mu_{\delta}$=-40 mas yr$^{-1}$  &$\mu_{\alpha}cos \delta$=+40 mas yr$^{-1}$ $\mu_{\delta}$=-20 mas yr$^{-1}$ &\textit{Z}\\ 
\hline
     0.61&  1.00&1.00& 21 - 22     \\
     0.35&  0.67&0.33& 22 - 23     \\
    -0.16& -2.00&0.00& 23 - 24    \\
\hline
\end{tabular}
\end{center}
\end{table*}  

\section{Results}
Most of these objects, except two bright objects and the faintest seven have been documented before in surveys - Moraux et al (2003) and Bihain et al
 (2006). We recovered 
all of these objects within our overlapping area, and none were rejected by our \textit{IZ} photometric selection.  
The objects we recovered were BRB 4, 8, 17, 13, 19, 21, 22, 27 and 28 and PLIZ 2, 3, 5, 6, 13, 14, 19, 20, 26, 28, 31, 34, 35 and 36. PLIZ 18, 27 and 39 were found to 
have no \textit{J} counterpart in our catalogues.
Of these objects, BRB 19 and PLIZ  14  and 26 met by our selection criteria on the \textit{Z}, \textit{Z}-\textit{J} CMD, however they 
 were too blue in their  \textit{Z}-\textit{J} colour for their place on the sequence.
Out of the remaining objects we find that we 
agree with the proper motion measurements as calculated by Bihain et al.(2006) for PLIZ 28,  which we believe is a member of the cluster. We agree with Bihain et al.(2006) over their candidates BRB 13 and BRB 19
that they are not proper motion members to the cluster, however  we disagree with their proper motion measurement for BRB 19.  
We also find that PLIZ 5 is a  non member to the cluster - ie its proper motion measurement 
is not within 14 mas yr$^{-1}$ of the cluster proper motion value. We find that PLIZ  14 and 26 are not  proper motion members to the cluster, 
as well as not having met our selection criteria. 
PLIZ 26 was found to have a proper motion measurement of 35.73$\pm$9.00, -25.83$\pm$6.96, which did not fall within 14 mas y$^{-1}$ of the cluster, and also missed 
the selection made with the wider circle (21 mas yr$^{-1}$) as well.
We find that PLIZ 19, 20, 34 and 36 are not proper motion members to the cluster. However this means we disagree
with Moraux et al. (2003), over their object PLIZ 20. They find a proper motion of 25.6$\pm$7.3, -44.7$\pm$7.4 mas yr$^{-1}$ for it. Our proper motion measurement is 0.88$\pm$ 15.86, -0.92$\pm$8.42 mas yr$^{-1}$. 
It is possible that this object has been adversely affected by
its position on the edge of one of the WFCAM chips, thus reducing the number of reference stars used to calculate its proper motion. 
 An alternative method of measuring the proper motion using all the objects on the same chip produced a measurement of 19.14$\pm$11.06,  -28.989$\pm$11.94 mas yr$^{-1}$. This value does meet our selection criteria, and has been previously accepted as a member. 
We suggest PLIZ 20 is likely to be a member because of this.

We find that PLIZ 2, 3, 6, 31 and 35 are all proper motion members to the cluster.
In addition to this, we find 2 brighter new candidate members to the cluster. These objects are bright enough to have appeared in previous 
surveys, and in the UKIDSS Galactic cluster survey (GCS). 
We also have 2 fainter new members to the cluster, and 5 objects selected using the \textit{ZJ} photometry only. All of the objects identified as cluster members in this work are presented in Table 4. 
\begin{table*}
\caption{}
\end{table*}
\begin{figure*}
\begin{center}
\scalebox{0.85}{\includegraphics[angle=180]{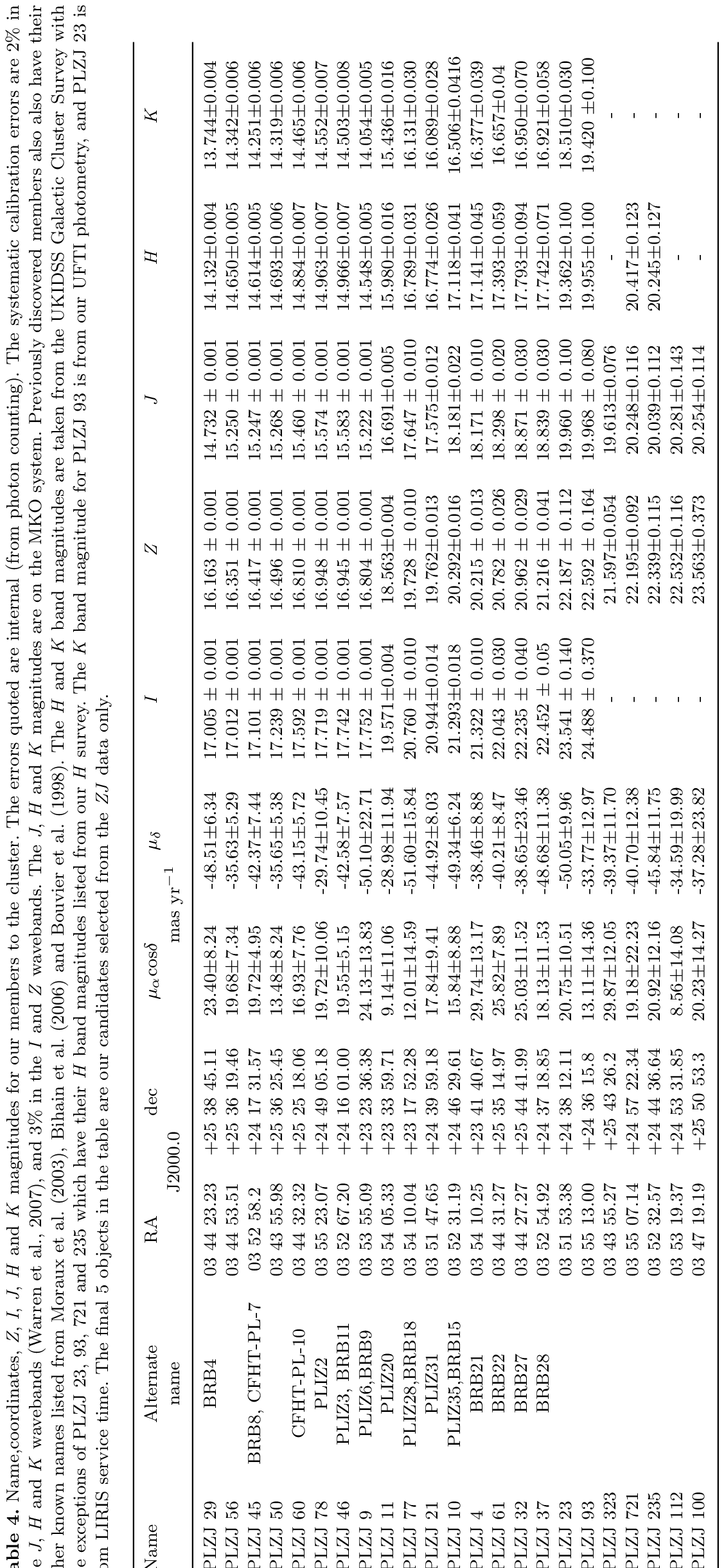}}
\end{center}
\end{figure*}
Two WFCAM tiles, 1 and 4, (see Figure 1) also had deep \textit{H} band photometry. 
These tiles were observed at the same time as the \textit{J} band imaging, and were observed under the same conditions,
 but with the exception that microstepping was not used.
These data were reduced using the same pipeline as the \textit{J} band data, but the photometry and object detection used a core radius of 2.5 
pixels in this case.
Fortunately these tiles also covered our faintest, previously undiscovered 
Pleiades 
candidates, PLZJ 23 and PLZJ 93, as well as two of the candidates selected from the \textit{ZJ} data only, PLZJ 721 and 235. 

The UKIDSS Galactic Cluster survey (GCS) has also covered the entire area at \textit{J}, \textit{H} and \textit{K}.  
The UKIDSS data are reduced using the same pipeline as the WFCAM data (see Dye et al, 2006 for details of the pipeline).

We also have used UKIRT service time
 to measure photometry for PLZJ 93 in the \textit{K} band. 
This observation was taken on 09/09/2006 in seeing of better than 1.1'' using the UKIRT Fast Track Imager (UFTI), with a five point dither pattern. 
The data were reduced using the ORAC-DR pipeline, and the photometry was
calibrated using UKIRT Faint Standard 115.

The \textit{K} band photometry for PLZJ 23 was obtained on the night of 05/03/2007 using the long slit intermediate resolution spectrograph (LIRIS) on the William 
Hershel Telescope in service time, using a nine point dither pattern in seeing of $\approx$ 0.9''.
The data were reduced using IRAF and astrometrically and photometrically calibrated using 2MASS.
The colour transforms presented in Carpenter, (2001) were used to calculate the \textit{K} band magnitude from the \textit{K}$_{S}$ magnitude.

Thus we have \textit{I}, \textit{Z}, \textit{J}, \textit{H} and \textit{K} band 
photometry for the majority of our Pleiades candidates.
However \textit{H} or \textit{K} band photometry is still needed for  PLZJ 323, 721, 235, 112 and 100, (see Table 4).
Figures 7 and 8 show the \textit{K}, \textit{J}-\textit{K} and \textit{H}, \textit{J}-\textit{H}, colour magnitude diagrams, 
together with the NEXTGEN (Baraffe et al, 1998) and DUSTY (Chabrier et al, 2000) 
models for the Pleiades age of 120 Myrs (Stauffer et al 1998).
The candidate members listed in Table 4 are also plotted in Figures 3 and 4 for clarity.
In both of these diagrams the M dwarf tail, the redward L dwarf sequence and the L to T blueward transition sequence are clear.  
The L-T transition sequence of course only has 
two objects plotted on it on Figure 7 as we have no \textit{K} band photometry  for the \textit{ZJ} candidates.  As expected the \textit{K},
 \textit{J}-\textit{K} diagram gives the best differentiation between the 
sequences.  The redward L sequence in this diagram agrees with that found by Lodieu et al,(2007b) derived from a much 
greater area of the Pleiades by the UKIDSS GCS.
The GCS is not sensitive enough to see the L-T blueward transition sequence however.
The \textit{K}, \textit{J}-\textit{K} diagram also shows the separation between single and binary dwarfs quite clearly.
 Note that the DUSTY theoretical track is too flat compared to our 
empirical sequence, see figures 7 and 8.

\begin{figure}
\begin{center}
\scalebox{0.35}{\includegraphics[angle=270]{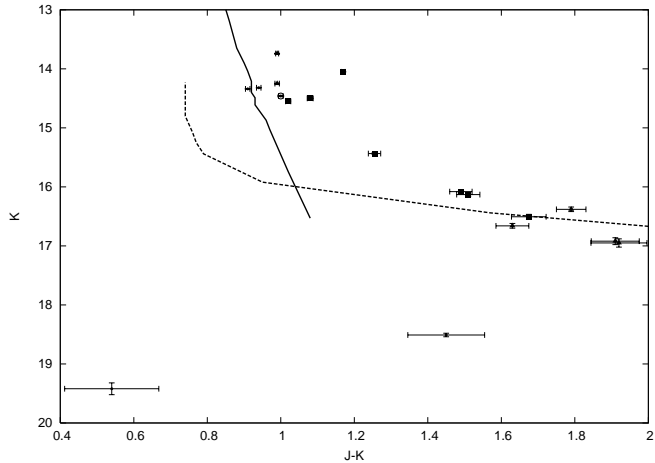}}
\caption{The \textit{K},\textit{J}-\textit{K} CMD for our candidate cluster members. The solid
line is the NEXTGEN model of Baraffe et al (1998), and the dotted line is the DUSTY model of Chabrier et al. (2000). 
The filled squares are the candidates identified by Moraux et al. (2003), the filled triangles are the candidates identified by 
Bihain et al. (2006), the object enclosed by the open circle is CFHT-PL-10 identified by Bouvier et al. (1998). The objects marked by 
small points are our new
candidate members.
One of our T dwarf candidates, PLZJ 93, is found to the bottom of the 
plot, with a \textit{J}-\textit{K} of $\approx$ 0.6. PLZJ 23 is also present with a \textit{J}-\textit{K} of 1.45.
}
\end{center}
\end{figure}
\begin{figure}
\begin{center}
\scalebox{0.35}{\includegraphics[angle=270]{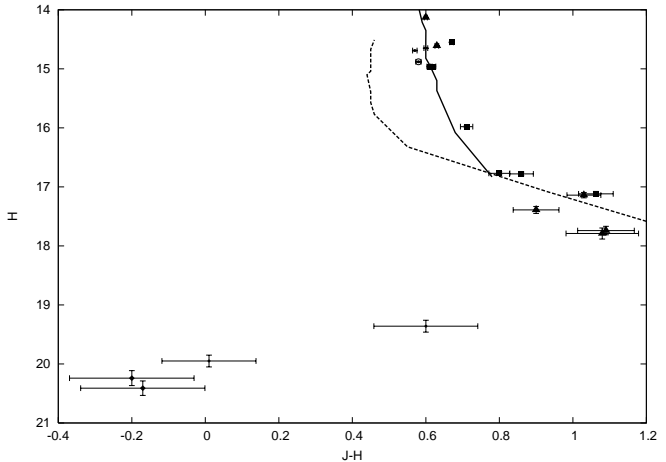}}
\caption{The \textit{H},\textit{J}-\textit{H} CMD for our candidate cluster members. 
The filled squares are the candidates identified by Moraux et al. (2003), the filled triangles are the candidates identified by 
Bihain et al. (2006), the object enclosed by the open circle is CFHT-PL-10 identified by Bouvier et al. (1998). The objects marked by small points are our new
candidate members.The filled diamonds are the two candidates with \textit{H} magnitudes 
selected from the \textit{ZJ} data only. The solid
line is the NEXTGEN model of Baraffe et al (1998), and the dotted line is the DUSTY model of Chabrier et al. (2000). 
}
\end{center}
\end{figure}

PLZJ 23, 93, 721 and 235 have \textit{J}-\textit{H} colours of 0.60, 0.00, 
-0.17 and -0.21 respectively.
Comparing these colours with the spectral type colour relations of field dwarfs described in Leggett et al. (2002), yields estimated 
spectral types of T1.5, T4.5, T6 and T6 respectively.  PLZJ 93 has \textit{J}-\textit{K}=0.60 which gives a spectral type of T3 (Leggett et al., 2002), 
which is consistent with the spectral type derived from the \textit{J}-\textit{H} colour (T4.5), within the errors.
We also can calculate a \textit{H}-\textit{K} colour for this dwarf of 0.6, however the \textit{H}-\textit{K} colour is not a good choice for 
spectral typing, for instance, \textit{H}-\textit{K}=0.6 covers a range of spectral types from L1 to T3 (Chiu et al., 2006). The \textit{Z}-\textit{J}
colour is also not a good choice of colour for measuring spectral types until the later T dwarfs ($>$T2)(Hawley et al., 2002). 
PLZJ 23 has \textit{J}-\textit{K}=1.45, which gives a spectral type of between L8 and T1. The \textit{H}-\textit{K} colour for this dwarf is 0.85.
We may thus assume that PLZJ 23 has a spectral type between L8 and T1.5, and likewise that PLZJ 93 has a spectral type of between T3 and T5 to take into account the 
photometric errors.
It should be noted 
that the Z band quoted in Hawley et al., (2002) is for the Sloan filter system, and so for this reason we have not chosen to use it to spectral type our objects.
We believe that the \textit{J}-\textit{H} colour gives the best estimate available to us of spectral types. Two of the three candidate members without \textit{H}
band photometry PLZJ 112 and 100  have fainter \textit{J} magnitudes than PLZJ 23 and 93, and so it is likely that they are also T dwarfs.
PLZJ 323 is brighter and is therefore probably a late L dwarf.
Indeed our faintest candidate at \textit{Z}, PLZJ 100, may be a very late T dwarf, however this assumption is made using its \textit{Z}-\textit{J} colour, 
which is very red.
Using \textit{J} magnitudes and the COND models of Baraffe et al. (2003) for 120 Myrs (the DUSTY models are no longer appropriate for calculating masses
for objects this faint in the Pleiades), we calculate masses of $\approx$ 11 M$_{\rm Jup}$ for PLZJ 23, 93, 323, 721, 235, 112 
and 100. More photometry in the \textit{H} and \textit{K} bands is clearly needed to improve and extend these estimates of the spectral types.

\section{Mass spectrum}
To calculate the mass spectrum, we first divided the sample into single dwarfs or single dwarfs with possible low mass companions and dwarfs that are
 close to 0.75 magnitudes above the single star sequence in the \textit{K}, \textit{J}-\textit{K} colour magnitude diagram.  The latter we assume to be equal mass binaries
 and count them as dwarfs with masses the same as a dwarf on the single dwarf sequence below them. From Figures 3, 4, 7 and 8
it can be seen that there are 2 such binaries all with \textit{J}-\textit{K} $\approx$ 1.
Dwarfs with \textit{J}-\textit{K} $<$1.2 are assigned masses using their H magnitudes and the NEXTGEN models (Baraffe et al. 1998).  For 
 1.2$<$\textit{J}-\textit{K}$<$2.0
we use the DUSTY models (Chabrier et al., 2000) and the \textit{J}-\textit{H} colour to assign a mass. Finally the T dwarf masses were 
calculated from their \textit{J}
 magnitudes and the COND models (Baraffe et al., 2003).
The masses were binned into three mass intervals, covering the low, medium and high mass ranges 
 and the numbers per bin are weighted by the probabilities of membership calculated
using the annulus, and the bin width has been taken into account. The candidate members with negative probabilities are obviously omitted from the mass spectrum.
The resultant mass spectrum is shown in Figure 9.  The errors are poissonian. Clearly the statistics are very poor, due to the small 
number of objects being dealt with. Using linear regression we have fitted our data  to the relationship dN/dM$\propto$M$^{-\alpha}$,  and calculate  
$\alpha$=0.35$\pm$0.31. 
This is lower but still in agreement with 
values presented in the literature (within 1$\sigma$), however the error on this value is large, and the statistics are poor due to the small numbers involved.
If we take into account the fact that the last mass bin is only 50\% complete (using Tables 1 and 4), then the lowest mass bin can be increased by 50\% to
 compensate. If we then fit these data, we derive a value for $\alpha$ of 0.62$\pm$0.14, which is in agreement with the literature. 
Alternatively, we can discount this final low mass bin as being incomplete and simply omit it from the fit. 
In this case we calculate a value for $\alpha$ of 0.86.
We have only displayed the mass spectrum for the cluster in the area and magnitude surveyed. This is 
to avoid trying to take into account biases caused by some areas being more studied than others, and also because we are only adding a maximum of 9 
objects to the mass spectrum, 7 of which have low probabilities of membership and small masses, and so are not likely to affect previous results a large amount.
  The mass spectrum appears to be rising towards the lowest masses, but this is not statistically 
significant due to the large error bars.

\begin{figure}
\begin{center}
\scalebox{0.35}{\includegraphics[angle=270]{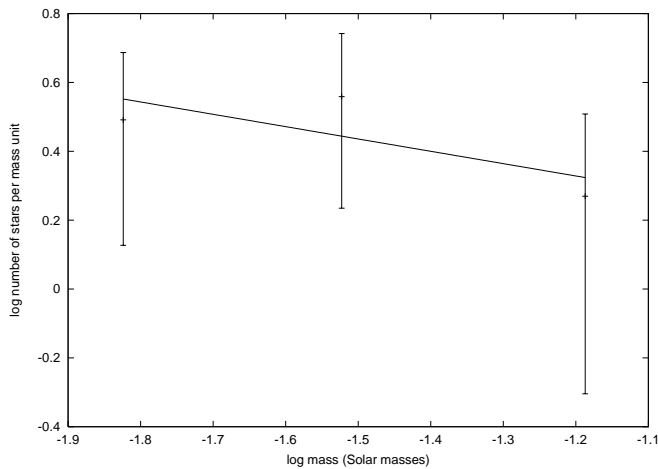}}
\caption{\label{mf}The mass spectrum for our Pleiades candidate members. The mass bin is in units of M$_{\odot}$. The solid line is the fit to the data,
($\alpha$=0.35$\pm$0.31). }
\end{center}
\end{figure}
\section{Conclusions}
We have confirmed a number of L dwarf candidates in the Pleiades.  However the main result in this paper is the discovery of seven L and  T dwarf 
Pleiads of masses $\approx$ 11 M$_{\rm Jup}$,  below the 13 M$_{\rm Jup}$ deuterium burning limit that is often used, somewhat artificially
 as the upper bound for planetary masses.  Further \textit{H} and \textit{K} band photometry, currently lacking for some of these candidates,
will improve confidence in their membership of the cluster.
 Planetary mass brown dwarfs have, of course, been claimed in the Orion nebula 
(Lucas \& Roche 2000) and in the $\sigma$-Ori cluster (Zapatero-Osorio et al., 2002).  These clusters both have very young ages and may also have a
 spread of ages (B\'ejar et al., 2001), making mass determinations somewhat uncertain. 
Lodieu et al. (2006, 2007a) have also found planetary mass brown dwarfs in the Upper Scorpius Association which has an age of 5 Myrs
 (Preibisch \& Zinnecker, 2002). At very young ages the theoretical models may have significant errors when used to assign masses (Baraffe et al., 2002).  
Our result is the
 first detection of planetary mass objects in a mature cluster whose age is well established. It strengthens the case that the star
 formation process can produce very low mass objects.
\section{Acknowledgements}
SLC, NL,and PDD acknowledge funding from PPARC.
We also acknowledge the Canadian Astronomy Data Centre, which is operated by the Dominion Astrophysical Observatory for the 
National Research Council of Canada's Herzberg Institute of Astrophysics.
This work has been based on observations obtained at the Canada-France-Hawaii Telescope (CFHT) which is operated by the National Research Council of Canada, the
Institut National des Sciences de l'Univers of the Centre National de la Recherche Scientifique of France, and the University of Hawaii. Observations were also made  at the
 United Kingdom Infrared Telescope, which is operated by the Joint Astronomy Centre on behalf of the U.K. Particle Physics and Astronomy Research Council.
This publication makes use of data products from the Two Micron All Sky Survey, which is a joint project of the University of Massachusetts and the Infrared Processing and Analysis Center/California Institute of Technology, funded by the National Aeronautics and Space Administration and the National Science Foundation.
This research has made use of NASA's Astrophysics Data System Bibliographic Services, the WHT service programme, and the UKIRT service programme.
We would like to thank the referee V.J.S. B\'ejar for his comments which have improved the paper.
\label{lastpage}
{}
\end{document}